\documentclass[aps,twocolumn,showpacs,preprintnumbers,amsmath,amssymb]{revtex4}

\usepackage{graphicx}% Include figure files
\usepackage{dcolumn}% Align table columns on decimal point
\usepackage{bm}% bold math
\usepackage{amssymb}
\usepackage{amsmath}
\begin{document}

%\preprint{APS/123-QED}

\title{A Novel Field Approach to 3D Gene Expression Pattern Characterization}

\author{L. da F. Costa}
 \email{luciano@if.sc.usp.br}
\author{B. Traven\c{c}olo}
 \email{bant@if.sc.usp.br}
\author{A. Azeredo}
\affiliation{Instituto de F\'{\i}sica de S\~{a}o Carlos,
Universidade de S\~{a}o Paulo, Av. Trabalhador S\~{a}o Carlense
400, Caixa Postal 369, CEP 13560-970, S\~{a}o Carlos, SP, Brazil}

\author{M. E. Beletti}
% \email{mebeletti@ufu.br}
\affiliation{Instituto de Ci\^{e}ncias Biom\'{e}dicas, Universidade
Federal de Uberl\^{a}ndia, Av. Par\'{a} 1720, CEP 38400-902, Uberl\^{a}ndia,
MG, Brazil}

\author{D. Rasskin-Gutman}
\author{G. Sternik}
% \altaffiliation[Also at ]{Salk Institute}
\author{J. C. I. Belmonte}
\author{M. Iba\~{n}es}
% \altaffiliation[Also at ]{Salk Institute}
\affiliation{Salk Institute, 10010N Torrey Pines Road, La Jolla,
USA CA 92037}

\author{G. B. M\"{u}ller}
\affiliation{Department of Zoology, University of Vienna,
Althaustrasse 14, A-1090 Vienna Austria.}

\date{\today}

\begin{abstract}
We present a vector field method for obtaining the spatial
organization of 3D patterns of gene expression based on gradients
and lines of force obtained by numerical integration. The
convergence of these lines of force in local maxima are identified
as centers of gene expression, providing a natural and powerful
framework to characterize the organization and dynamics of
biological structures. We apply this novel methodology to analyze
the expression patterns of light chain myosin II protein linked to
enhanced green fluorescent protein (EGFP) during zebrafish heart
formation.
\end{abstract}

%\pacs{Valid PACS appear here}

\maketitle

Animal development involves synchronized gene activation modulated
by environmental influences \cite{Carroll:2001,Gilbert:2003}. Far
from being uniform, such a gene expression gives rise to
structured spatial and temporal patterns of varying protein
concentration. Recent advances in biochemical and imaging methods
have paved the way to obtaining 3D reconstructions of spatial gene
activation \cite{Streicher:2000} which can be analyzed in order to
better understand the intricate mechanisms governing tissue, organ
and member formation \cite{Gilbert:2003}. Among the several
currently available methodologies allowing characterization of 3D
gene expression, special attention has been given to EGFP
(Enhanced Green Fluorescence Protein). The EGFP is used as a
marker. Its expression is controlled by the promoter of the gene
of interest creating a fluorescent fusion protein that maintains
the normal functions and localization of the wild type protein.
This methodology can be used to demonstrate gene activity in
intact cells and organisms, while taking into account the fact
that the host protein is continuously synthesized, degraded, and
suffering alterations within cells
\cite{Tisen:1998,Patterson:2003}.  As such a type of gene
expression data becomes available, it is important to identify and
develop mathematical methologies for measuring and modeling
spatial gene activation. In addition to traditional approaches
(e.g.  density or dispersion estimation), it is important to
consider more sophisticated methods capable of addressing more
directly aspects related to the dynamics of the involved
biological processes, such as cell communication and migration
\cite{Schock:2002,Kuure:2000}, which play an important role during
both embryonic development and pathological processes.

In this article we characterize the spatial organization of gene
expression patterns in order to assess the geometrical basis of some
dynamical processes during morphogenesis. To this end, we compute a
``gene expression landscape'' as a scalar field $\omega = g(x, y, z)$,
where $\omega$ is interpreted as the amount of expression of the
protein in the spatial position $(x, y, z)$. The same approach can be
used to model and predict the dissemination of cell signalling or
other influence factors emmanating from the cell under analysis which,
combined with the possibility of adopting varying values of the
parameters affecting the field (e.g. the dielectric constant), defines
a truly general framework for expressing field influences.  In analogy
with the potential dynamics of dissipative systems, we obtain the
spatial trajectories (lines of force) corresponding to maximizing the
gradient of gene expression.  Such trajectories tend to converge to
local peaks of activity, defining gene expression centers.  It is
proposed in this article that the distribution of such centers provide
a natural framework for characterizing and analyzing the spatial
interactions between the involved developmental rudiments.  The
potential of such a methodology is illustrated with respect to the
analysis of zebrafish heart formation from 3D gene expression data.

Zebrafish embryos have been widely used in order to study heart
formation, due to their transparency and its partial independence
from the cardiovascular system. For vertebrates, the heart is the
first organ that forms and starts operating~\cite{Stainier:2001}.
Constrictions and bending (folding) are key elements in the early
morphogenetic shaping of the heart tube.  The spatial gene
expression data considered in this work was acquired through the
observation of 42-hour post-fertilization transgenic zebrafish
embryos expressing EGFP specific for heart mesoderm myosin light
chain (mlc2a-EGFP)~\cite{Raya:2003}. The zebrafish embryos were
anesthetized and kept fixed, and live-images of the heart were
taken at ambient room temperature. The image recordings were made
using a Nikon Eclipse TE300 inverted microscope using 20x/0.75 NA
magnification. The microscope is coupled to a Bio-Rad Radiance MP
2100 scanning multiphoton confocal system (Cambridge,MA) with a
two-photon Tsunami laser (Spectra Physics, CA). The GFP was
excited with the two-photon laser, at 900 nm. The total dataset is
composed of 110 confocal sections.

All the 110 confocal slices were combined so as to obtain the
three-dimensional volume of the heart, from which the gene
expression landscape was computed as described above.  It is
interesting to note that this scalar field can be visualized with
direct volume rendering algorithms (DVR)~\cite{Schroeder:1996}. In
order to minimize the spatial quantization noise implied by
digital image representation, gaussian smoothing was applied over
the gene concentration data. This is done through the discrete
convolution of a three-dimensional Gaussian kernel {\it k(x,y,z)}
with the scalar field {\it w}, as expressed in Eq.
(\ref{Eq_GaussSmooth})

\begin{eqnarray} \label{Eq_GaussSmooth}
w(x,y,z)*k(x,y,z) &=& \sum_{i,j,k}w(i,j,k) \nonumber \\ &\times&
k((x-i),(y-j),(z-k))
\end{eqnarray}

\begin{figure*}
\includegraphics[scale=1]{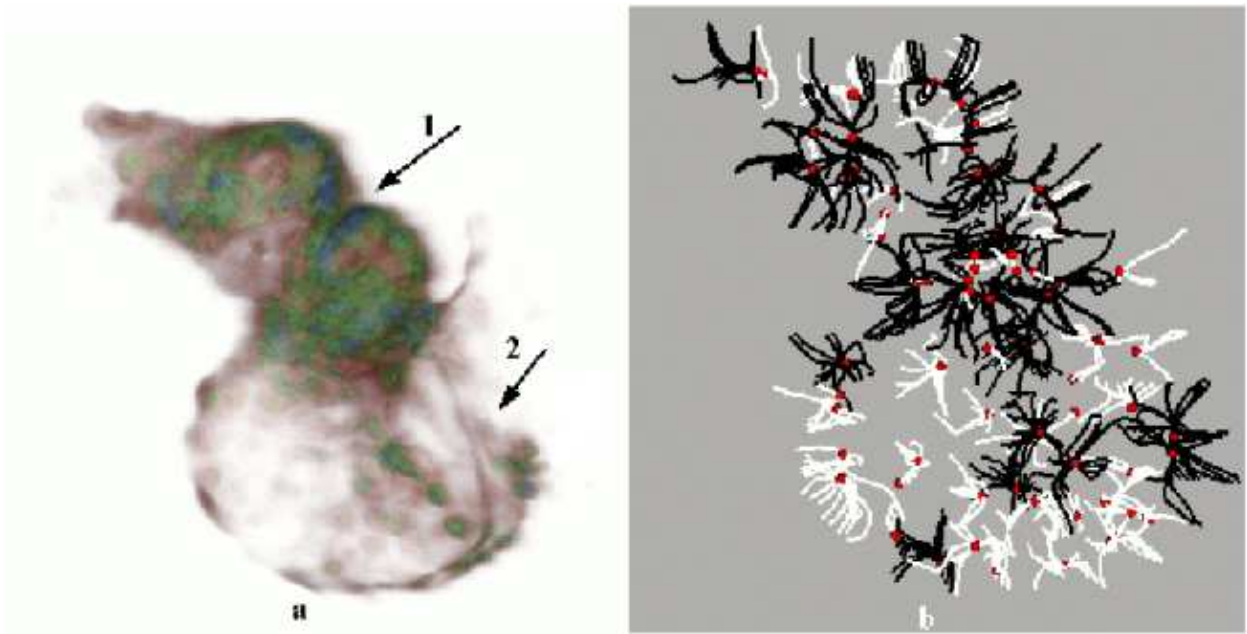}
\caption{\label{Fig_Cluster}(a) Visualization of the smoothed and
reconstructed gene activity pattern of mlc2a during zebrafish
heart formation. The inflow pole is on the upper left. Arrows~1
and 2 indicates the constriction$/$bending regions. The respective
lines of force are shown in (b), segregated into black and white
as described in the text.}
\end{figure*}

The smoothed reconstruction of the 3D gene activity pattern is
shown in Figure~\ref{Fig_Cluster}a. The gradient of this scalar
field was estimated by using the enhanced finite differences
scheme described in \cite{Zucker:1981}, by convolving the gene
expression concentration with three-dimensional masks. Next, we
compute the lines of force by calculating the trajectories that
maximize the gradient starting form arbitrary spatial positions
sampled as points uniformly distributed through spheres centered
at the three-dimensional volume.

The considered lines of force would correspond, for instance, to
the putative path (set of 3D coordinates) followed by an object at
position $\vec{r}=(x,y,z)$ with gradient dissipative dynamics:

\begin{equation}
\frac{\partial \vec{r}}{\partial
t}=\vec{\nabla}\{\omega(x,y,z)*k(x,y,z)\}\,,
\end{equation}

standard numerical integration was used in order to estimate such
lines of force, which are illustrated in
Figure~\ref{Fig_Cluster}b. The sampling criteria removed the lines
whose scalar value of its end point were less than 10 (from a
range of 0 to 255), eliminating those that do not reach the
regions where mlc2a was being expressed. Small and too long
trajectories were also removed, because they were influenced by
noises. As expected, these lines converge to local maxima of the
scalar gene expression field, which could be considered as
\emph{gene expression centers}. In analogy to graph theory, the
total number of sampled lines of force converging to a specific
center is referred to as the center \emph{degree}.  A total of 734
lines and 89 centers were obtained for the considered 3D gene
expression data.

\begin{figure}[t]
\includegraphics[scale=1]{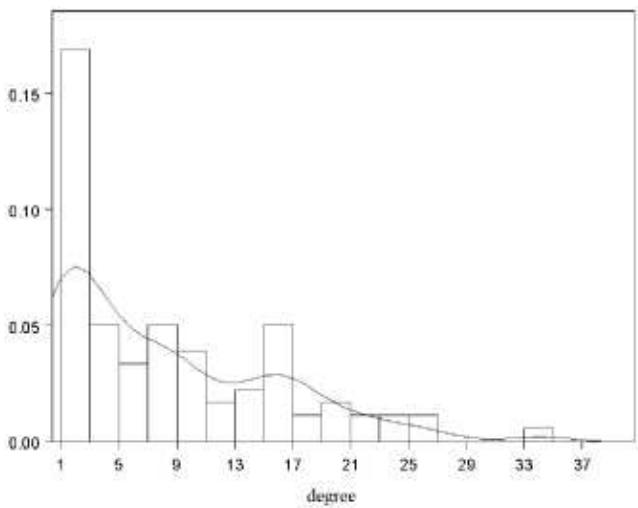}
\caption{\label{Fig_Hist} Relative frequency histogram of the
distribution of centers degree}
\end{figure}

Figure ~\ref{Fig_Cluster}b shows the sampled lines of force
obtained by using the above described methodology, drawn in black
or white according to thresholding criteria: the lines
corresponding to gene expression activity centers with degrees
smaller than 14 have been marked in white. Such threshold value
was defined with basis on the relative frequency histogram of the
distribution of centers degree, showed in Figure~\ref{Fig_Hist}.
It can be seen from Figure~\ref{Fig_Cluster}b that the genic
activity centers exhibiting higher numbers of converging lines of
force (marked black) tend to concentrate along the regions
subjected to the constriction and folding implied by the heart
formation dynamics (marked by arrow~$1$ in Figure
~\ref{Fig_Cluster}a) as well as the sinus venosus (marked by
arrow~$2$ in Figure ~\ref{Fig_Cluster}a).  The following
biological interpretation are suggested in order to account for
such result.

The heart forms from a tube of epimyocardial cells that express,
among other genes, mlc2a. This gene is expressed uniformly
throughout the heart, with the possibility of a weaker expression
in the inflow pole, i.e. the region of the venous sinus and the
atrium (Figure~\ref{Fig_Cluster}a). It is suggested here that the
distribution of active cells could be determined by a gene
activity field in such a way that the higher degree activity
centers positively regulates the activation patterns of
surrounding cells. The line of force pattern indicates that the
expression of these cells coincides with morphogenetic events of
heart formation, in particular the characteristic constrictions
and bendings of the heart tube at the atrio-ventricular and the
ventriculo-bulbar borders (arrows in Figure~\ref{Fig_Cluster}a),
which are sites composed by high degree activity centers,
involving cells more actively producing mlc2a. This process might
be affected by the differential distribution of gene activation
centers, as indicated by the respective numbers of lines of force
which tended to be smoother at these locations.

While such hypotheses can only be verified through further
experimental investigations, a novel methodology for 3D gene
activity characterization has been shown to provide a natural and
effective means for quantifying the spacial interactions between
the biological structures involved in gene expression.  Unlike
differential measurements such as gradients or divergent
magnitudes, the estimation of the lines of force and activity
centers are integral features, indicating spatial interactions
over substantial distances. It is expected that the proposed
framework will prove to be useful in a number of other gene
expression investigations, paving the way to a more objective
understanding of the dynamics governing animal development and its
pathologies.

\begin{acknowledgments}

The authors thank HFSP RGP39/2002 for funding this project.
A. Azeredo are grateful to FAPESP (02/09149-2), and B. Traven\c{c}olo
is grateful to CAPES and FAPESP (03/13072-8) for financial
support. M. Iba\~nes is partially supported by the Fulbright Program
and Generalitat of Catalunya. Luciano da F. Costa is grateful to
FAPESP (proc. 99/12765-2) and CNPq (proc. 301422/92-3) for financial
support.

\end{acknowledgments}

%\newpage %Just because of unusual number of tables stacked at end
\bibliography{pre3}% Produces the bibliography via BibTeX.

\end{document}